\documentclass[11pt]{article}
\usepackage{amsmath,amssymb}
\usepackage{epsfig}
\def\dref#1{(\ref{#1})}

\begin{document}

\vspace*{1\baselineskip} \begin{center}\Large\bf Network
synchronizability analysis: the theory of subgraphs and
complementary graphs \footnote {\small
 This work was supported by the
 National
 Science Foundation of
 China under grant 60674093 and the  City
University  of  Hong  Kong  under  the  Research  Enhancement Scheme
and  SRG  grant  7002134.}
 \end{center}
\vspace*{1\baselineskip}

\centerline {Zhisheng Duan$^1$,   ~ Chao Liu$^1$, ~and ~Guanrong
Chen$^{1,2}$ }

\vspace*{0.5\baselineskip}
\begin{center}
 $^1$State Key Laboratory for Turbulence and Complex Systems,
Department of Mechanics and Aerospace Engineering, College of
Engineering, Peking University, Beijing 100871, P. R. China\\
$^2$ Department of Electronic Engineering, City University of Hong
Kong, Hong Kong
 \\ Email:
       duanzs@pku.edu.cn,  {\it Fax}: (8610)62765037
\end{center}
\vskip 0.5cm

 {\bf Abstract.} \,\,  In this paper, subgraphs and complementary
graphs  are used to  analyze the network synchronizability. Some
sharp  and attainable bounds are provided for the eigenratio of the
network structural matrix, which characterizes the network
synchronizability, especially when the network's corresponding graph
has cycles, chains, bipartite graphs or product graphs as its
subgraphs.


\section{Introduction}

Synchronization of complex networks has attracted increasing
attention from many scientists, for its important applications in
various areas of physical and biological sciences \cite{bar02,
bel05, don05, hol02, hong04, wang06, wat98}. Synchronization is a
ubiquitous  phenomenon especially in social and biological networks
where, quite often, it is desirable, e.g., in the consensus of
multi-agent activities, while in some other cases it is undesirable,
e.g., in traffic congestion \cite{don05, mot05, nis06, nis03, sor07,
ste07, wang02}. The connection structure of a network plays an
important role in its synchronization. Of particular interest in
this concern is how the synchronizability depends on various
structural parameters of the network, such as average distance,
 clustering coefficient, coupling strength, degree distribution and weight distribution,
 among others. Some interesting  results have been established for such
 important
problems based on the notions of master stability function and
synchronized region \cite{bar02, boc06, koc05, luw04, lu04,
mot05,pec98,zhou06}. Some relationships between synchronizability
and structural parameters of complex networks have also been
reported, e.g., smaller average network distance does not
necessarily mean better synchronizability \cite{nis03}, therefore
the betweenness centrality was proposed as a good indicator for
synchronizability \cite{hong04}, and two networks with the same
degree sequence can have different synchronizabilities \cite{wu05}.
 Some optimizing rules were established for the network
synchronizability \cite{don05}. Moreover, how the network
statistical properties influence the spectrum of the structural
matrix was analyzed and reported in \cite{atay06}. And the effects
of graph operations on the network synchronizability were studied in
\cite{atay05}. Last but not least, complementary graphs were used to
analyze the network synchronizability in
 \cite{duan07}. Motivated by all these research works, this paper attempts to
  further establish some sharp upper bounds for the network synchronizability based on the
  theory of subgraphs and
  complementary graphs.

Consider a dynamical network consisting of $N$ coupled identical
nodes, with each node being an $n$-dimensional dynamical system,
described by
\begin{equation}\label{n1}
\dot{x}_i=f(x_i)-c\sum_{j=1}^Na_{ij}H(x_j),\;i=1,2,\cdots,N,
\end{equation}
where $x_i=(x_{i1},x_{i2},\cdots,x_{in})\in \mathbb{R}^n$ is the
state vector of node $i$, $f(\cdot):\mathbb{R}^n\rightarrow
\mathbb{R}^n$ is a smooth vector-valued function, constant $c>0$
represents the coupling strength, $H(\cdot):\mathbb{R}^n\rightarrow
\mathbb{R}^n$ is called the inner linking function, and
$A=(a_{ij})_{N\times N}$ is called the outer coupling matrix or
structural matrix, which represents the coupling configuration of
the entire network. This paper only considers the case that  the
network is diffusively connected, i.e., $A$ is irreducible and its
entries satisfy $ a_{ii}=-\sum_{j=1,j\neq
i}^Na_{ij},\;i=1,2,\cdots,N. $ Further, suppose that, if there is an
edge between node $i$ and node $j$, then $a_{ij}=a_{ji}=-1$, i.e.,
$A$ is a Laplacian matrix corresponding to network (\ref{n1}).
Therefore, $0$ is an eigenvalue of $A$ with multiplicity 1, and all
the other eigenvalues of $A$ are strictly positive, which are
denoted by
 \begin{equation}\label{f1}
0=\lambda_1<\lambda_2\leq\lambda_3\leq\cdots\leq\lambda_N=\lambda_{max}.
\end{equation}
 The dynamical network \dref{n1} is said to achieve (asymptotical)
synchronization if
$x_1(t)\rightarrow x_2(t)\rightarrow\cdots\rightarrow
x_N(t)\rightarrow s(t),\;\textrm{as}\; t\rightarrow \infty,$
where, because of the diffusive coupling configuration,  the
synchronous state $s(t)\in \mathbb{R}^n$ is a solution of an
individual node, i.e., $\dot{s}(t)=f(s(t))$.

It is well known \cite{bar02} that the eigenratio
$r(A)=\frac{\lambda_2}{\lambda_N}$ of network structural matrix $A$
characterizes the network synchronizability: the larger the $r(A)$,
the better the synchronizability. In this paper, $r(A)$ will be used
as a synchronizability index for the networks with structural matrix
$A$.  The relationships between $r(A)$ and network structural
characteristics such as average distance, node betweenness, degree
distribution, clustering coefficient, etc. have been  studied
\cite{duan07, atay06, wu05, zhao06}. A well-known upper bound for
the eigenratio $r(A)$ in graph theory  is $r(A)\leq
\frac{d_{min}}{d_{max}}$, where $d_{min}$ and $d_{max}$ denote the
smallest and largest degrees of the corresponding graph to the
network \cite{wu05}. To the best of our knowledge, besides this
coarse bound, there are no sharp upper bounds available in the
literature today. Some sharp upper bounds for $r(A)$ are derived in
this paper by using graph-theoretical methods.

The rest of this paper is organized as follows. In Section 2, some
necessary preliminaries in graph theory are provided. In Section 3,
for a given graph $G$, some upper bounds are derived for the
eigenratio $r(G)$ when $G$ has cycles as its subgraphs.  In Section
4, some  lower bounds for the largest eigenvalue of a graph are
presented when $G$ has other subgraphs, such as chains, bipartite
graphs and product graphs. The paper is concluded by the last
section.

Throughout this paper, for any given undirected graph $G$,
eigenvalues of $G$ mean eigenvalues of its corresponding Laplacian
matrix. Notations for graphs and their corresponding Laplacian
matrices are not differentiated, and networks and their
corresponding graphs are not distinguished, unless otherwise
indicated.

\section{Preliminaries}

For a given graph $G$, let ${\cal V}$ and ${\cal E}$ denote the sets
of nodes and edges of $G$, respectively.  A graph $G_1$ is called an
induced subgraph of $G$, if the node set ${\cal V}_1$ of $G_1$ is a
subset of ${\cal V}$ and the edges of $G_1$ are all edges among
nodes ${\cal V}_1$ in ${\cal E}$. The complementary graph  of $G$,
denoted by $G^c$, is the graph containing all the nodes of $G$ and
all the edges that are not in $G$.
 In this paper, subgraphs and
complementary graphs are used to discuss the network
synchronizability. For this purpose, the following lemmas
 are  needed.

{\bf Lemma 1}\,\cite{mer94, mer98}\, For any given
 graph $G$ of size $N$, its nonzero
  eigenvalues indexed as in
  (\ref{f1}) grow monotonically with the
 number of added edges; that is, for any added edge $e$,
 $\lambda_i(G+e)\geq \lambda_i(G)$, $i=1, \cdots, N$.

 {\bf Lemma 2}\,\cite{mer94}\, For any given connected
  graph $G$ of size $N$, its largest eigenvalue $\lambda_N$ satisfies
  $\lambda_N\geq d_{max} +1$, with equality if and only if $d_{max}=N-1$,
where $d_{max}$ is the maximum degree of $G$.

{\bf Lemma 3}\,\cite{god01, wu05}\, For any cycle $C_N$ at $N$
($\geq 4$) nodes, its eigenvalues are  given by $\mu_1, \cdots,
\mu_N$ (not necessarily ordered) with $\mu_1=0$ and
$$\mu_{k+1}= 3- \frac {\sin (\frac{3k\pi}{N})}{\sin(\frac{k\pi}{N})}, \,\, k=1,\cdots, N-1.$$

By Lemma 3, one knows that,
 for any cycle $C_N$ at $N$
($\geq 4$) nodes, if $N$ is even, its largest eigenvalue is 4; if
$N$ is odd, its largest eigenvalue is $ \lambda_N(C_N)= 3- \frac
{\sin (\frac{3(N-1)\pi}{2N})}{\sin(\frac{(N-1)\pi}{2N})}.$

Obviously, for cycles with odd numbers of nodes, the largest
eigenvalue converges to 4 as the number of its lengths  tends to
$+\infty$. For eigenvalues of graphs and complementary graphs, the
following lemma is useful (see \cite{mer94, mer98} and references
therein).

 {\bf Lemma 4}\,\, For any given graph $G$ of size $N$, the following statements
 hold:

(i) \, $\lambda_N(G)$, the largest eigenvalue of $G$, satisfies
$\lambda_N(G)\leq N.$

(ii) \, $\lambda_N(G)= N$ if and only if $G^c$ is disconnected.

(iii)\, If $G^c$ is disconnected and  has (exactly) $q$ connected
components, then the multiplicity of $\lambda_N(G)=N$  is $q-1$.

(iv) \, $\lambda_i(G^c)+\lambda_{N-i+2}(G)=N,\quad 2\leq i\leq N$.

For a given graph, generally its largest eigenvalue is easier to
compute than the smallest nonzero eigenvalue. Hence, by Lemma 4, one
can obtain the smallest nonzero eigenvalue by computing the largest
eigenvalue of its complementary graph.

{\bf Corollary 1}\,\, For any given graph $G$ of size $N$, if its
second smallest eigenvalue equals its smallest node degree, i.e.,
$\lambda_{2}(G)=d_{min}(G)$,  then $G$ or $G^c$ is disconnected; if
$\lambda_{2}(G)>d_{min}(G)$, then $G$ is a complete graph; if both
$G$ and $G^c$ are connected, then $\lambda_2(G)<d_{min}(G).$

{\bf Proof}\,\, If $G^c$ is connected, then
$\lambda_{max}(G^c)=N-\lambda_{2}(G)=N-d_{min}(G_1)=d_{max}(G^c)+1$.
By Lemma 2, $d_{max}(G^c)=N-1$, so $G$ is disconnected. Further, if
$\lambda_{2}(G)>d_{min}(G)$, by Lemma 4, $\lambda_N(G^c)<
d_{max}+1$. Combining with Lemma 2, $G^c$ can only have isolated
nodes, i.e., $G$ is a complete graph. The third statement holds
obviously. \hfill $\Box$

 Combining  Lemmas 1 and 2 with Corollary 1,  one can easily get
the following result.

 {\bf Corollary 2}\,\, For any given connected graph $G$ and every
 its induced subgraph $G_1$, one has $\lambda_{max}(G)\geq
 \lambda_{max}(G_1)$,  so the synchronizability index of $G$ satisfies
  $r(G)< \frac{d_{min}(G)}{\lambda_{max}(G_1)};$ if both $G$ and $G^c$ are connected, then
$r(G)< \frac{d_{min}}{d_{max}+1}.$

 Since subgraphs have less nodes, this corollary is useful when a
 graph $G$ contains some canonical graphs whose largest eigenvalues can be easily obtained as
  subgraphs (see the section below for further discussion).

 {\bf Corollary 3}\,\,
 For a given graph $G$, if the largest eigenvalue of $G^c$ is
 $\lambda_{max}=d_{max}(G^c)+\alpha$, then $\lambda_2(G)=d_{min}(G)+1-\alpha$. Consequently,
 the synchronizability index of $G$ satisfies
 $r(G)= \frac{d_{min}(G)+1 -\alpha }{\lambda_{max}(G)}
 \leq \frac{d_{min}(G)+1-\alpha}{d_{max}(G)+1}.$

By Lemma 2, generally $\alpha \geq 1$, so the bound in Corollary 3
is better than the one in Corollary 2.

\section{Graphs having cycles as subgraphs}

 In this section, an even cycle means a cycle with an even number of nodes and an odd cycle means
 a cycle with an odd number of nodes. By the discussion in the above section, one can get the following results.

{\bf Lemma 5}\,\, For any even cycle, $-2$ is an eigenvalue of its
adjacency matrix. For any odd cycle with $n_1$ nodes, $-1+\frac{\sin
(\frac{3(n_1-1)\pi}{2n_1})}{\sin(\frac{(n_1-1)\pi}{2n_1})}$ is an
eigenvalue of its adjacency matrix.


{\bf Proof}\,\, Lemma 3 directly leads to the result. \hfill $\Box$

 {\bf
Theorem 1}\,\, For any given graph $G$, suppose $G_1$ is its induced
subgraph composing of all nodes of $G$ with the maximum  degree
$d_{max}(G)$, and ${\cal G}_1$ is the induced subgraph of $G^c$
composing of all nodes of $G^c$ with the maximum  degree
$d_{max}(G^c)$. Then, the following statements hold:

(i) \,\, If both  $G_1$ and ${\cal G}_1$ have  even cycles as
induced subgraphs, then $\lambda_{max}(G)\geq d_{max}(G)+2$ and
$\lambda_{max}( G^c)\geq d_{max}( G^c)+2$. Consequently, the
synchronizability  index of $G$ satisfies $r(G)\leq
\frac{d_{min}(G)-1}{d_{max}(G)+2}$.

(ii) \,\, If both  $G_1$ and ${\cal G}_1$ have odd cycles as induced
subgraphs, and if the longest odd cycle of $G_1$ has $n_1$ nodes and
the longest odd cycle of ${\cal G}_1$ has $n_2$ nodes, then
$\lambda_{max}(G)\geq d_{max}(G)+1-\frac{\sin
(\frac{3(n_1-1)\pi}{2n_1})}{\sin(\frac{(n_1-1)\pi}{2n_1})}$ and
$\lambda_{max}(G^c)\geq d_{max}(G^c)+1-\frac {\sin
(\frac{3(n_2-1)\pi}{2n_2})}{\sin(\frac{(n_2-1)\pi}{2n_2})}$.
Consequently, the synchronizability  index of $G$ satisfies $r(G)
\leq \frac{d_{min}(G)+\frac {\sin
(\frac{3(n_2-1)\pi}{2n_2})}{\sin(\frac{(n_2-1)\pi}{2n_2})}}{d_{max}(G)+1-\frac
{\sin (\frac{3(n_1-1)\pi}{2n_1})}{\sin(\frac{(n_1-1)\pi}{2n_1})}}$.

(iii) \,\, If $G_1$ has an even cycle as an induced subgraph,  and
${\cal G}_1$ has odd cycles as induced subgraphs with the longest
odd cycle having $n_2$ nodes, then $\lambda_{max}(G)\geq
d_{max}(G)+2$ and $\lambda_{max}(G^c)\geq d_{max}(G^c)+1-\frac {\sin
(\frac{3(n_2-1)\pi}{2n_2})}{\sin(\frac{(n_2-1)\pi}{2n_2})}$.
Consequently, the synchronizability  index of $G$ satisfies \\$r(G)
\leq \frac{d_{min}(G)+\frac {\sin
(\frac{3(n_2-1)\pi}{2n_2})}{\sin(\frac{(n_2-1)\pi}{2n_2})}}{d_{max}(G)+2}$.

(iv) \,\, If  $G_1$ has odd cycles as induced subgraphs with the
longest odd cycle having $n_1$ nodes and ${\cal G}_1$ has an even
cycle as an induced subgraph, then $\lambda_{max}(G)\geq
d_{max}(G)+1-\frac{\sin
(\frac{3(n_1-1)\pi}{2n_1})}{\sin(\frac{(n_1-1)\pi}{2n_1})}$ and
$\lambda_{max}(G^c)\geq d_{max}( G^c)+2$. Consequently, the
synchronizability  index  of $G$ satisfies\\ $r(G) \leq
\frac{d_{min}(G)-1}{d_{max}(G)+1-\frac {\sin
(\frac{3(n_1-1)\pi}{2n_1})}{\sin(\frac{(n_1-1)\pi}{2n_1})}}$.

 {\bf Proof}\,\, (i)\,\,  By Lemmas 3 and 5,
for any even cycle, its largest eigenvalue is 4 and -2 is an
eigenvalue of its adjacency matrix. Let $L_1$ be the sub-matrix of
the Laplacian matrix of $G$ related to all the nodes in $G_1$, and
$A(G_1)$ be the adjacency matrix of $G_1$. Then, $A(G_1)$ contains a
sub-matrix which is the adjacency matrix of the corresponding even
cycle. Therefore, one has
$$(d_{max}(G)+2)I-L_1=2I+A(G_1)\not > 0.$$
In fact, the sub-matrix of  $2I+A(G_1)$ corresponding to the even
cycle has a zero eigenvalue. This implies that the largest
eigenvalue of $G_1$ is larger than or equal to $d_{max}+2$. Then, by
Lemma 1, $\lambda_{max}(G)\geq d_{max}+2$. Similarly,
$\lambda_{max}(G^c)\geq d_{max}+2$.

 Further, suppose the number of nodes of $G$ is $N$.
 Then, the smallest nonzero eigenvalue of $G$ is
 $\lambda_2=N-\lambda_{max}(G^c)$ and the minimum node degree of $G$ is $d_{min}(G)=N-1-d_{max}(G^c)$.
If $\lambda_{max}(G^c)\geq d_{max}(G^c)+2$, then $\lambda_2\leq
N-d_{max}(G^c)-2=d_{min}(G)-1$. Therefore, Theorem 1 (i) obviously
holds.

(ii)\,\,  Follows easily from  (i) and Lemma 5.

(iii) and (iv) can be similarly proved. \hfill $\Box$

{\bf Remark 1}\,\, For a given graph $G$, clearly, the bounds given
in Theorem 1 for the synchronizability index $r(G)$ are better than
the general bound $r(G)\leq \frac{d_{min}}{d_{max}}$. And the
existence of cycles can be easily tested by drawing graphs. Since
the networks with good synchronizabilities always have homogeneous
degree distributions \cite{don05, hong04}, Theorem 1 is especially
useful for  estimating the synchronizabilities of homogeneous
networks.

{\bf Remark 2}\,\, The smallest even cycle is cycle $C_4$, and its
complementary graph $C_4^c$ has two separated edges. $C_4$ and
$C_4^c$ are very important in graph theory \cite{mer98}. A graph has
a $C_4$ as an induced subgraph if and only if $G^c$ has $C^c_4$ as
an induced subgraph.

{\bf Remark 3}\,\, The cycle $C_3$ is the smallest cycle. If a graph
$G$ contains only $C_3$ cycles, its largest eigenvalue satisfies
$\lambda_{max}(G)\geq d_{max}+1.$

 {\bf Example 1}\,\, Consider cycle $C_5$ in Fig. 1. Its
 complementary graph is also $C_5$ (see Fig. 2). Testing the
 eigenvalues of $C_5$ and its synchronizability, one finds that they
 attain the exact  bounds in Theorem 1 (ii).

\begin{center}
 \unitlength=1cm
 \qquad \hbox{\hspace*{0.1cm}  \epsfxsize5cm \epsfysize5cm
\epsffile{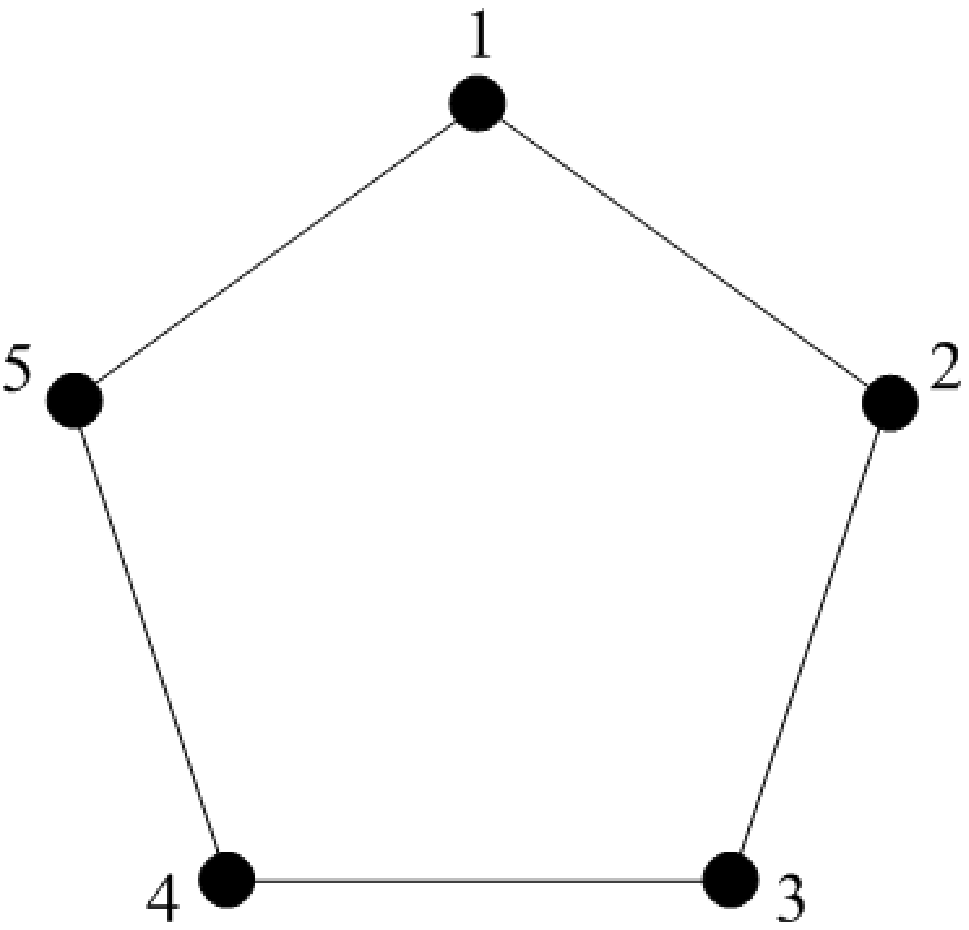}\qquad \epsfxsize4.5cm \epsfysize4.5cm
\epsffile{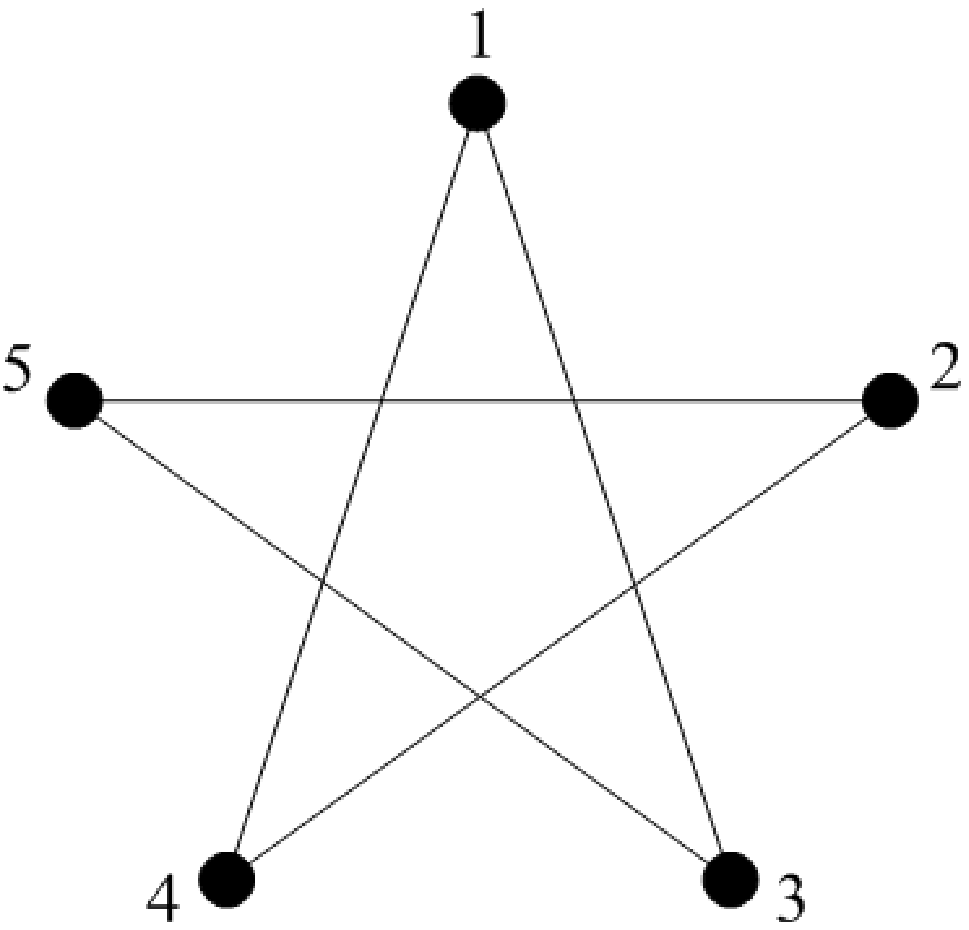} $\Leftrightarrow$  \epsfxsize4.5cm
\epsfysize4.5cm \epsffile{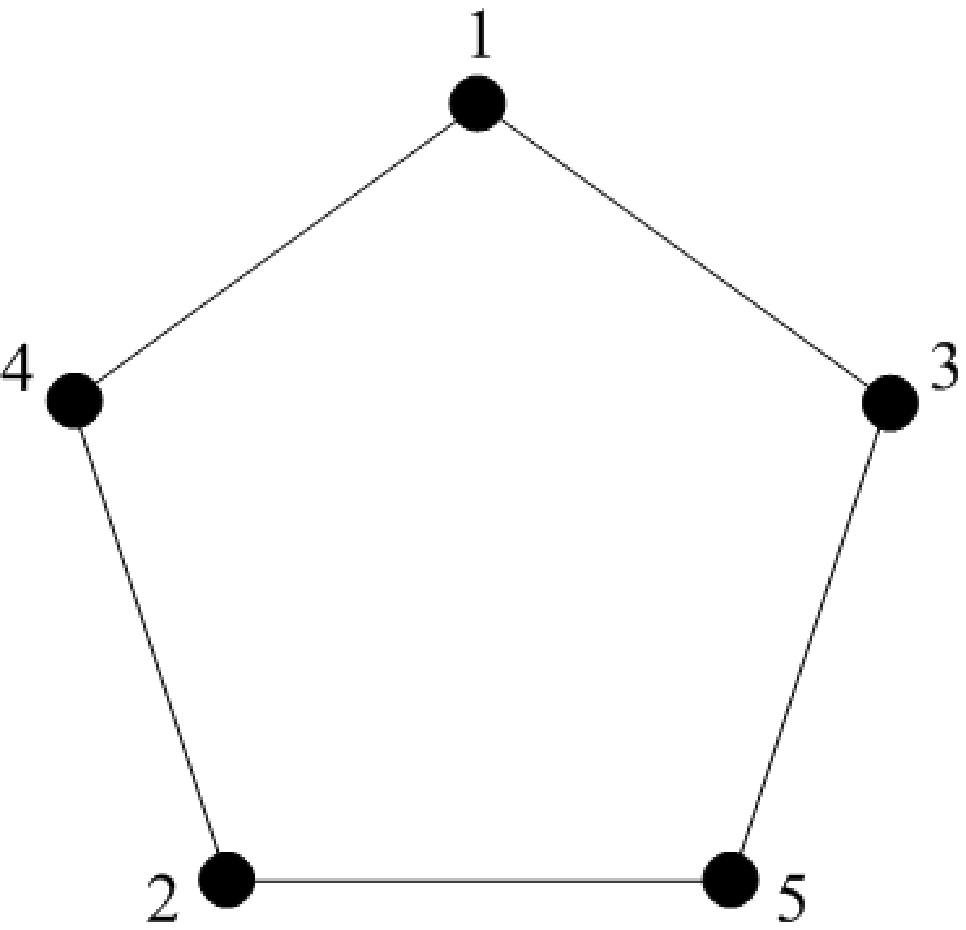}
 }
\end{center}
\vskip -0.3cm \quad\qquad Fig. 1 \,\, Graph $C_5$ \qquad\qquad\qquad
\qquad\qquad\qquad Fig. 2 \,\, Graph $C_5^c$

{\bf Example 2}\,\, Consider graph $\Gamma_1$ in Fig. 3. Its
 complementary graph is $C_6$ (see Fig. 4). Testing the
 eigenvalues of $\Gamma_1$ and its synchronizability, one finds that they
 attain the exact bounds in Theorem 1 (i).

\begin{center}
 \unitlength=1cm
 \qquad \hbox{\hspace*{0.1cm}  \epsfxsize5cm \epsfysize5cm
\epsffile{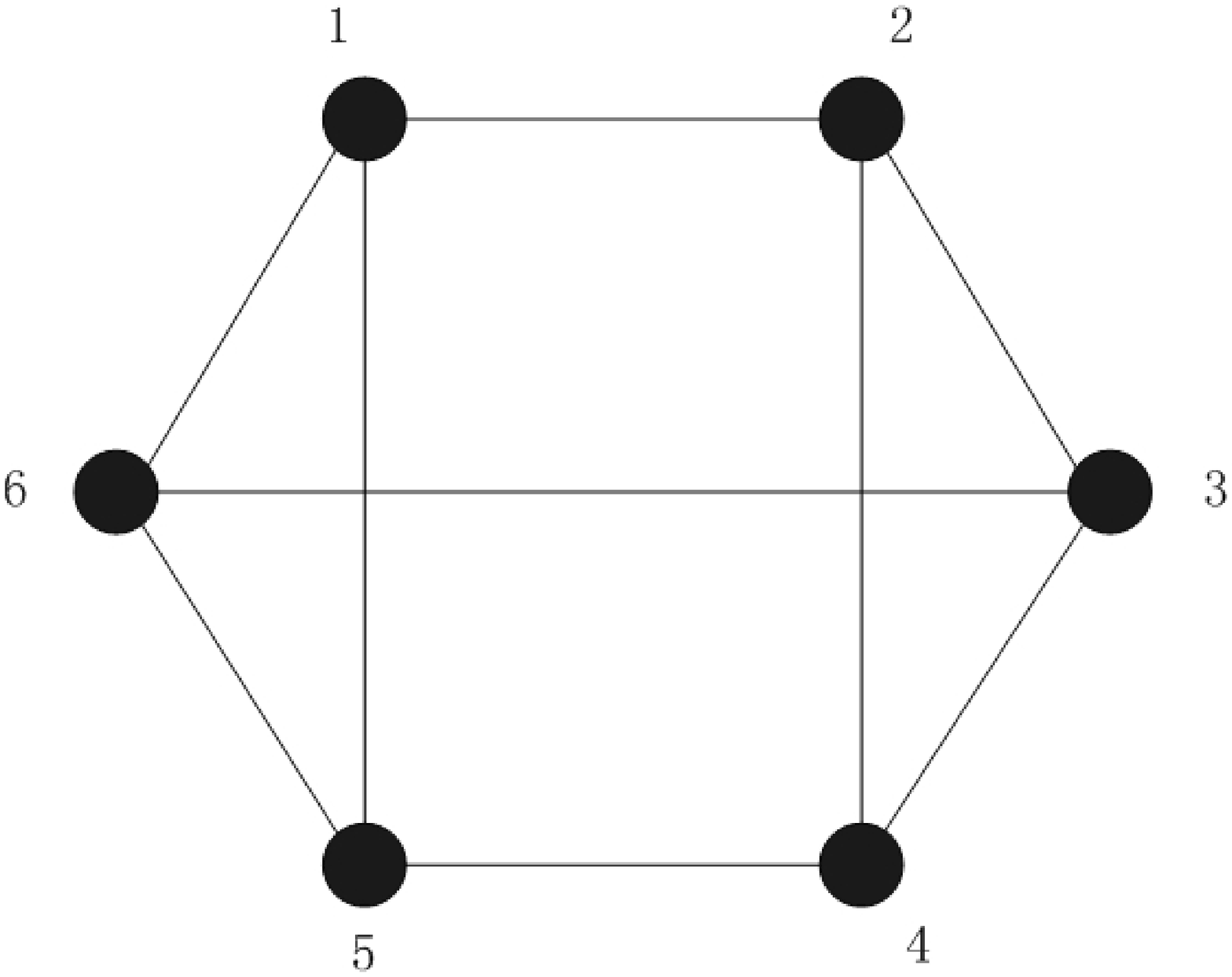}\qquad \epsfxsize4.5cm \epsfysize4.5cm
\epsffile{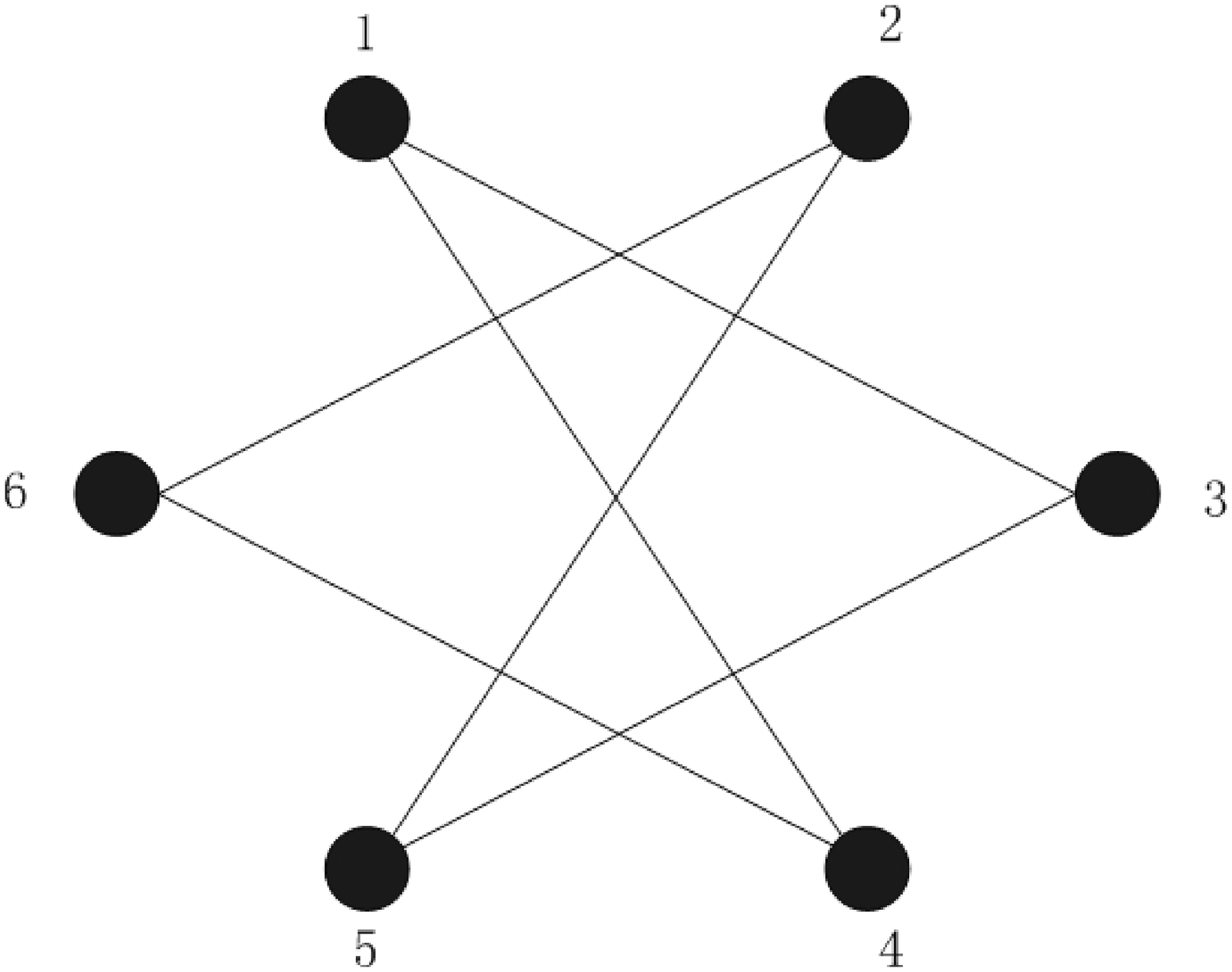} $\Leftrightarrow$  \epsfxsize5cm \epsfysize5cm
\epsffile{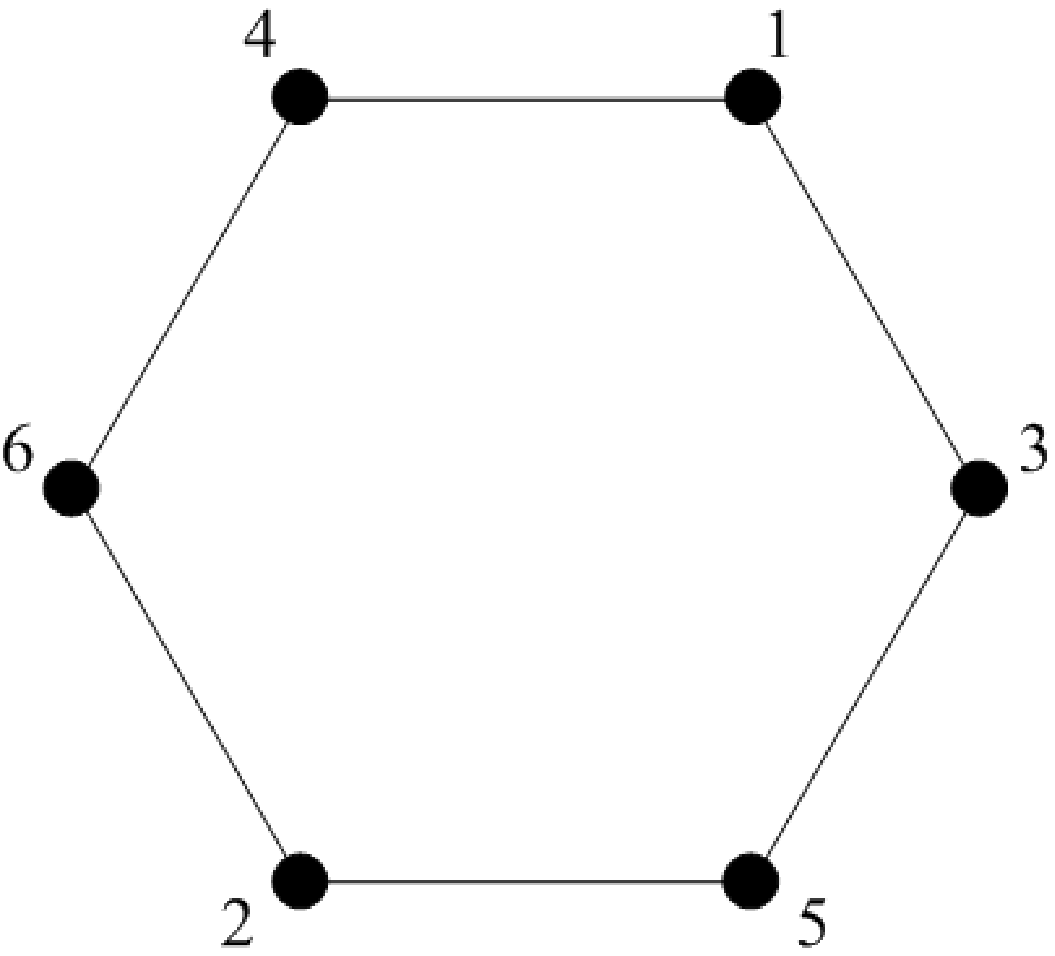}
 }
\end{center}
\vskip -0.3cm \qquad Fig. 3 \,\, Graph $\Gamma_1$ \qquad\qquad\qquad
\qquad\qquad\qquad Fig. 4 \,\, Graph $\Gamma_1^c=C_6$

\section{Applications of other subgraphs }

According to the discussion in the above section, cycles are very
important for  estimating the network synchronizability. In this
section, consider the estimation of the largest eigenvalue of a
given graph by its subgraphs. First, the following lemma for
adjacency matrices is needed.

 {\bf Lemma 6}\,\, \cite{nik06}\, Given a  graph $G$, the smallest
eigenvalue of its adjacency matrix $A(G)$ satisfies
$\lambda_{min}(A(G))\leq d_{max}(G)-\lambda_{max}(G).$

By this lemma, one can get the following result.

 {\bf Theorem 2}\,\, For a given graph $G$, let $H$
be a subgraph of $G$ containing all nodes of $G$ with the same node
degree $d$. Let $H_1$ be a subgraph of $H$.  Then, the largest
eigenvalue of $G$ satisfies $\lambda_{max}(G)\geq
d+\lambda_{max}(H_1)-d_{max}(H_1).$

{\bf Proof} \,\, It follows from Lemma 6 and the proof of Theorem 1.
 \hfill $\Box$

{\bf Remark 4}\,\, In the above section, cycles as subgraphs have
the same node degrees. So the smallest eigenvalues of the adjacency
matrices of cycles can be exactly computed. In Theorem 2, for a
general subgraph $H_1$, its node degrees are not necessarily the
same. In this case, Lemma 6 is very useful for  estimating  the
largest eigenvalues.

 In what follows, consider some canonical subgraphs which can be used to
estimate the largest eigenvalues of the original graphs. All the
results can be combined with the results in the above section to
obtain upper bounds for the network synchronizability.

\subsection{Graphs having chains as subgraphs}

A chain is a nearest-neighbor coupled graph without an edge between
the first and the last nodes. For a given chain $P_N$ with $N$
nodes,  it is well known that its largest and smallest nonzero
eigenvalues are $2(1+\cos(\frac{\pi}{N}))$ and
$2(1-\cos(\frac{\pi}{N}))$, respectively \cite{god01, mer94, wu05}.
Obviously, the largest eigenvalue of a chain converges to $4$ as its
length tends to $+\infty$.  According to the above discussion, one
can establish the following result.

{\bf Theorem 3}\,\, For a given graph $G$, let $G_1$ be a subgraph
of $G$ containing all nodes of $G$ with the maximum degree. Suppose
$G_1$ contains a chain $P_k$ with $k$ nodes as its subgraph. Then,
the largest eigenvalue of $G$ satisfies $\lambda_{max}(G)\geq
d_{max}+2\cos(\frac{\pi}{N}).$

{\bf Proof}\,\, Theorems 1 and 2 directly lead to the conclusion.
\hfill $\Box$

Note that chain $P_4$ is a very important graph in graph theory
\cite{mer98}. The complementary graph of $P_4$ is still a $P_4$
graph. Therefore, a graph $G$ has $P_4$ as a subgraph if and only if
$G^c$ has $P_4$ as a subgraph.

\subsection{Graphs having bipartite graphs as subgraphs}

Let $G_1({\cal V}_1, {\cal E}_1)$ and $G_2({\cal V}_2, {\cal E}_2)$
be two graphs on disjoint sets of $n_1$ and $n_2$ nodes,
respectively, where ${\cal V}_i$ and ${\cal E}_i$ are the
corresponding sets of nodes and edges.  Their disjoint union
$G_1+G_2$ is the graph $G_1+G_2=({\cal V}_1\bigcup {\cal V}_2, {\cal
E}_1\bigcup {\cal E}_2).$  Then, a bipartite graph $G_1*G_2$
generated by $G_1$ and $G_2$ is a new graph composing of $G_1+G_2$
and the new edges connecting each node of $G_1$ to every node of
$G_2$, as detailed in \cite{atay05, mer98, pan02}. It is well known
that the largest eigenvalue of the bipartite graph $G_1*G_2$ is
$\lambda_{max}(G_1*G_2)=n_1+n_2,$ i.e., the sum of the numbers of
nodes of $G_1$ and $G_2$. In fact, the complementary graphs of
bipartite graphs are always disconnected.

By Lemma 6, for a bipartite graph $G_1*G_2$, as discussed above, one
can has $\lambda_{min}(A(G_1*G_2))\leq d_{max}(G_1*G_2)-n_1-n_2.$

{\bf Theorem 4}\,\, For a given graph $G$, let $H$ be a subgraph of
$G$ containing all nodes of $G$ with the same degree $d$. Suppose
$H$ contains a bipartite subgraph $H_1*H_2$, and the numbers of
nodes of $H_1$ and $H_2$ are $n_1$ and $n_2$, respectively. Then,
the largest eigenvalue of $G$ satisfies $\lambda_{max}(G)\geq
d+n_1+n_2-d_{max}(H_1*H_2).$

{\bf Proof}\,\, This is a direct result of Theorem 2 for bipartite
subgraphs. \hfill $\Box$


{\bf Example 3}\,\, Consider graph $\Gamma_2$ in Fig. 5. Obviously,
$\Gamma_2$ has a bipartite graph $H$ as its subgraph which is
composed of all nodes with degree 6 from $\Gamma_2$. The largest
eigenvalue of this bipartite graph is 8. So, by Theorem 4,
$\lambda_{max}(\Gamma_2)\geq 9.$ On the other hand, the maximum
degree of the complementary graph $\Gamma_2^c$ of $\Gamma_2$ is 8.
And the nodes with degree 8 in $\Gamma_2^c$ form a cycle $C_4$. By
Theorem 1, the smallest nonzero eigenvalue of $\Gamma_2$ satisfies
$\lambda_2(\Gamma_2)\leq d_{min}(\Gamma_2)-1=2.$ So,
$r(\Gamma_2)\leq \frac{2}{9}.$ Simply computing the Laplacian
eigenvalues of $\Gamma_2$, one has $\lambda_2=1.7251$ and
$\lambda_{max}=9.2749.$ Consequently, $r(\Gamma_2)=
\frac{1.7251}{9.2749}\approx 0.176.$ Therefore, the theorems in this
paper successfully give the upper integer of the largest eigenvalue
and the lower integer of the smallest nonzero eigenvalue of
$\Gamma_2$.

\begin{center}
 \unitlength=1cm
 \qquad \hbox{\hspace*{0.1cm}  \epsfxsize7.5cm \epsfysize6cm
\epsffile{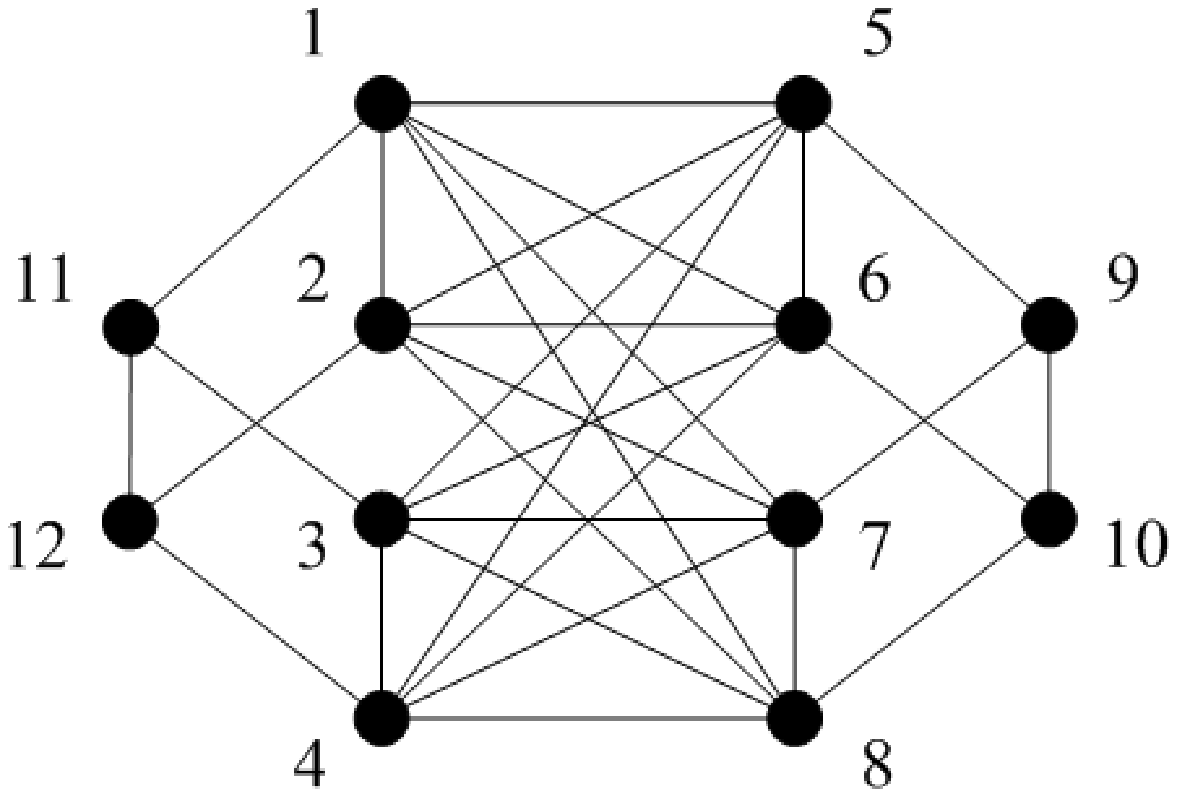}
 }
\end{center}
\vskip -0.3cm \centerline{\quad  Fig. 5 \,\, Graph $\Gamma_2$ }

\subsection{Graphs having product graphs as subgraphs}

Let $G_1({\cal V}_1, {\cal E}_1)$ and $G_2({\cal V}_2, {\cal E}_2)$
be two nonempty graphs. Their product graph $G_1\times G_2$ is a
graph with node set ${\cal V}_1 \times {\cal V}_2$, and $(x_1,
x_2)(y_1, y_1)$ is an edge in ${\cal E}(G_1\times G_2)$ if and only
if either $x_2=y_2$ with $x_1y_1\in {\cal E}(G_1)$ or $x_1=y_1$ with
$x_2y_2\in {\cal E}(G_2)$. One may view $G_1\times G_2$ as the graph
obtained from $G_1$ by replacing each of its nodes with a copy of
$G_2$ and each of its edges with the number of nodes of $G_2$ edges
joining the corresponding nodes of $G_2$ in two copies,
 as detailed in \cite{atay05, mer98}.

It is well known that the largest eigenvalue of the product graph
$G_1\times G_2$ is $\lambda_{max}(G_1\times
G_2)=\lambda_{max}(G_1)+\lambda_{max}(G_2).$

{\bf Theorem  5}\,\, For a given graph $G$, let $H$ be a subgraph of
$G$ containing all nodes of $G$ with the same node degree $d$.
Suppose $H$ contains a product graph $H_1\times H_2$ as its
subgraph. Then, the largest eigenvalue of $G$ satisfies
$\lambda_{max}(G)\geq
d+\lambda_{max}(H_1)+\lambda_{max}(H_2)-d_{max}(H_1\times H_2).$

{\bf Example 4}\,\, Consider graph $\Gamma_3$ in Fig. 6. Obviously,
all nodes of $\Gamma_3$ have degree 4. And $\Gamma_3$ has a product
graph $C_4\times P_3$ (nodes 1 to 12) as its subgraph, where $P_3$
denotes a chain with three nodes.  The largest eigenvalue of this
product subgraph is 7. So, by Theorem 5 or Lemma 1,
$\lambda_{max}(\Gamma_3)\geq 7.$ On the other hand,  the
complementary graph $\Gamma_3^c$ of $\Gamma_3$ has a bipartite graph
as its subgraph which is composed of nodes 1 to 4 and nodes 9 to 12.
By Theorem 4, similarly to Example 3, the largest eigenvalue of
$\Gamma_3^c$ satisfies $\lambda(\Gamma_3^c) \geq
d_{max}(\Gamma_3^c)+3$. Thus, by Corollary 3, the smallest nonzero
eigenvalue of $\Gamma_3$ satisfies $\lambda_2(\Gamma_3)\leq
d_{min}(\Gamma_2)-2=2.$ So, $r(\Gamma_3)\leq \frac{2}{7}.$ Simply
computing the Laplacian eigenvalues of $\Gamma_3$, one has
$\lambda_2=1.2679$ and $\lambda_{max}=7.4142.$ Consequently,
$r(\Gamma_2)= \frac{1.2679}{7.4142}\approx 0.171.$ Similarly to
Example 3, the corresponding theorems proved in this paper
successfully give the upper integer of the largest eigenvalue and
the lower integer of the smallest nonzero eigenvalue of $\Gamma_3$.

\begin{center}
 \unitlength=1cm
 \qquad \hbox{\hspace*{0.1cm}  \epsfxsize7.5cm \epsfysize7.5cm
\epsffile{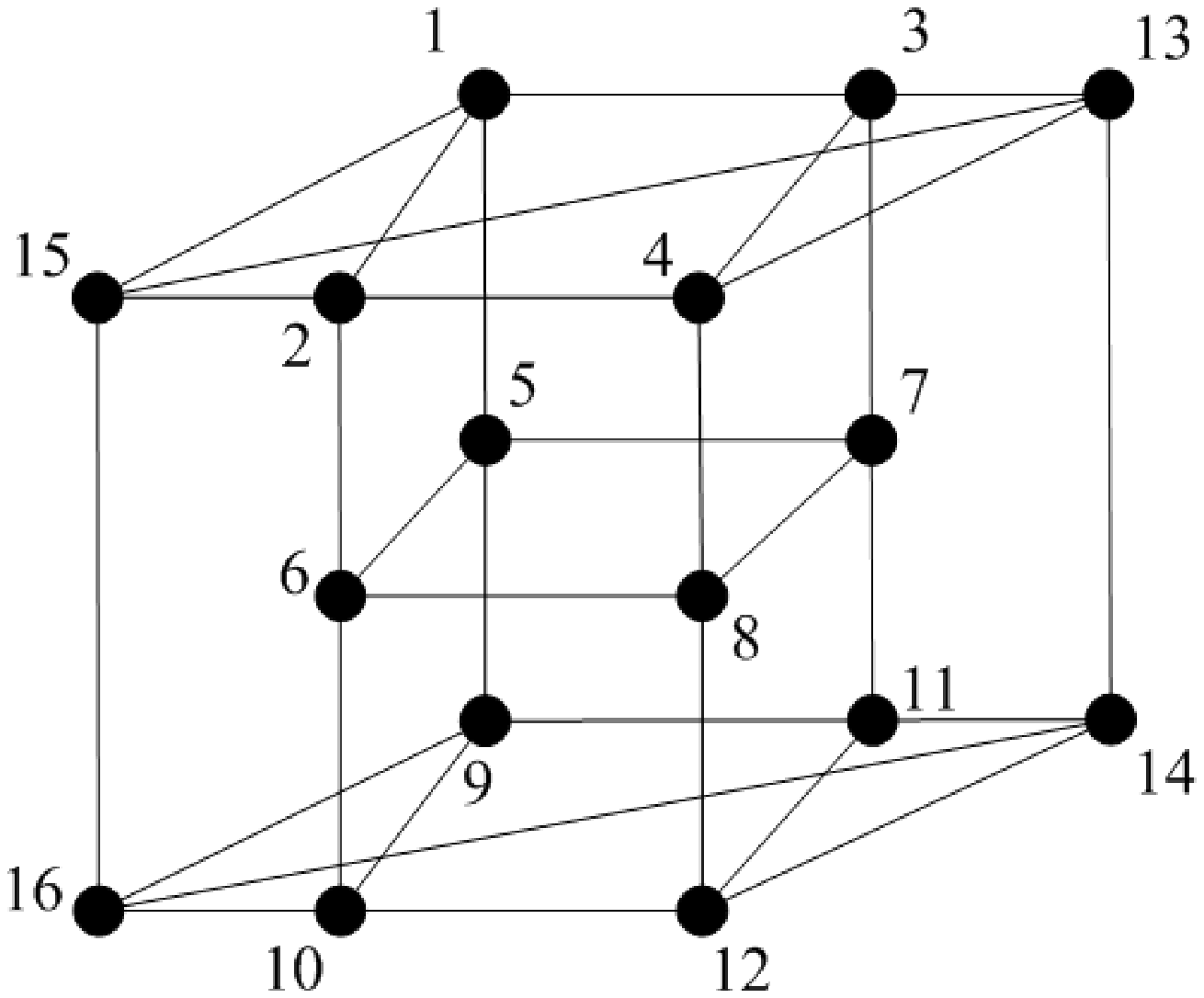}
 }
\end{center}
\vskip -0.3cm \centerline{  Fig. 6 \,\, Graph $\Gamma_3$ }

\subsection{The maximum disconnected subgraph}
Given a graph $G$ of size $N$, suppose $G_1$ is an induced subgraph
of $G$ of size $n_1$, and $G_1$ is disconnected. $G_1$ is called a
maximum disconnected subgraph, if the node number of any other
disconnected subgraph of $G$ is less than or equal to $n_1$.

{\bf Theorem  5}\,\, For a given connected graph $G$ of size $N$, if
the node number of its maximum disconnected subgraph is $n_1$,  then
the smallest nonzero eigenvalue of $G$ satisfies $\lambda_2 \leq
N-n_1$. Consequently, $r(G)\leq \frac{N-n_1}{d_{max}(G)+1}.$

{\bf  Proof}\,\, Suppose $G_1$ is a maximum disconnected subgraph of
size $n_1$. Then, by Lemma 4, $\lambda_{max}(G_1^c)\geq n_1$.
Further, $\lambda_{max}(G^c)\geq n_1$, so $\lambda_{2}(G)\leq N-
n_1$. \hfill $\Box$

{\bf Example 5}\,\, Consider graph $\Gamma_4$ in Fig. 7. By deleting
node 3 or 6, one can verify that the node number of its maximum
disconnected subgraph is 7. So, $\lambda_2(\Gamma_4)\leq 1$.
Combining Lemma 2 and Theorem 6, one has $r(\Gamma_4) <
\frac{1}{5}.$ It is well known that graphs as in Fig. 7 have large
node and edge betweenness centralities therefore have bad
synchronizabilities. Theorem 6, based on the theory of subgraphs and
complementary graphs, gives an explanation why such graphs indeed
have bad synchronizabilities.

\begin{center}
 \unitlength=1cm
 \qquad \hbox{\hspace*{0.1cm}  \epsfxsize5.5cm \epsfysize4.5cm
\epsffile{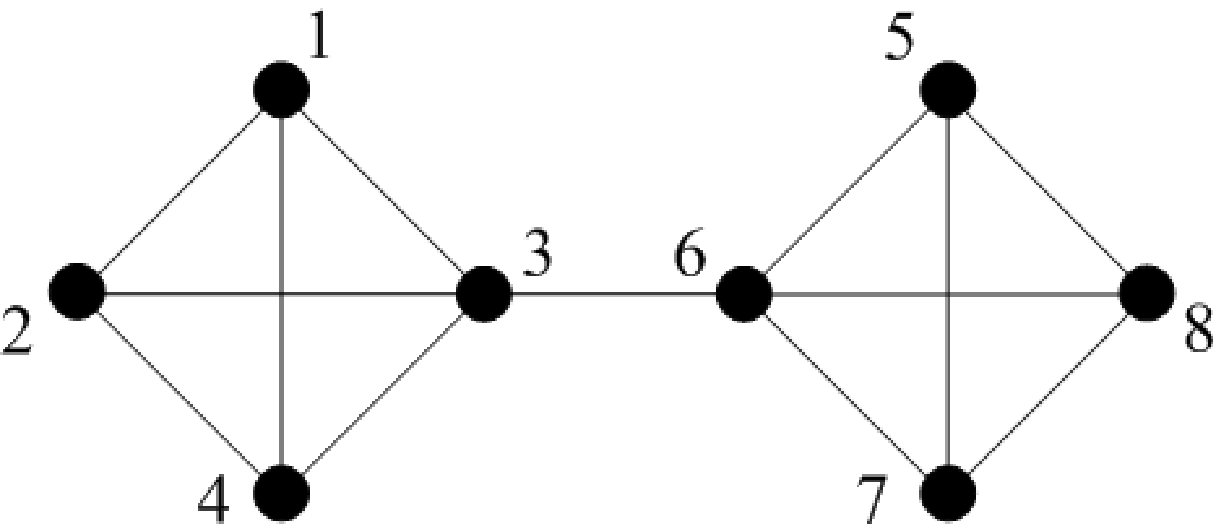}
 }
\end{center}
\vskip -0.3cm \centerline{  Fig. 7 \,\, Graph $\Gamma_4$ }

{\bf Remark 5}\,\,  From Examples 3 to 5, it can be seen that
actually one does not need to draw out the complementary graphs; the
properties of complementary graphs can be analyzed using the
original graphs.

\section{Conclusion}

In this paper, some relatively tight upper bounds have been obtained
for the network synchronizability based on the theory of subgraphs
and complementary graphs. Especially, some sharp bounds are given
for the eigenratioes and the largest eigenvalues when a graph has
some canonical subgraphs, such as cycles, chains, bipartite graphs
and product graphs.  Considering that the networks with good
synchronizabilities typically  have homogeneous degree
distributions, the results obtained in this paper
 are particularly  useful for estimating the synchronizabilities of homogeneous
 networks. And for a
given network, its corresponding graph as well as its complementary
graph and subgraphs can be easily obtained. This paper shows that
better understanding and helpful manipulation of subgraphs and
complementary graphs are very useful for enhancing the network
synchronizability. Therefore, the graph-theoretical method provided
in this paper is deemed important in the study of network
synchronization problems.

 \hspace*{34pt}

\end{document}